\documentclass[10pt]{article}

\usepackage{amsxtra,amssymb,amsthm,amsmath,latexsym}

\usepackage{graphicx}
\newtheorem{thm}{Theorem}[section]

\newcommand{\R}{{\mathbb R}}
\newcommand{\N}{\mathcal{ N}}
\newcommand{\Es}{\mathcal{E}}
\newcommand{\og}{\omega}
\newcommand{\ep}{\epsilon}
\newcommand{\al}{\alpha}

\newcommand{\dl}{{\delta}}
\newcommand{\bee}{\begin{equation*}}
\newcommand{\eee}{\end{equation*}}
\newcommand{\be}{\begin{equation}}
\newcommand{\ee}{\end{equation}}
\newcommand{\pn}{\par\noindent}
\title{Creating materials with a desired refraction coefficient}
\author{A G Ramm\\
\small Department of Mathematics\\[-0.8ex]
\small Kansas State University, Manhattan, KS 66506-2602, USA\\[-0.8ex]
\small \texttt{ramm@math.ksu.edu}\\
}

\begin{document}
 \date{}
\maketitle
\begin{abstract}
A method is given for creating material with a desired refraction
coefficient. The method consists of embedding into a material with
known refraction coefficient many small particles of size $a$. The
number of particles per unit volume around any point is prescribed,
the distance between neighboring particles is
$O(a^{\frac{2-\kappa}{3}})$ as $a\to 0$, $0<\kappa<1$ is a fixed
parameter. The total number of the embedded particle is
$O(a^{\kappa-2})$. The physical properties of the particles are
described by the boundary impedance $\zeta_m$ of the $m-th$
particle, $\zeta_m=O(a^{-\kappa})$ as $a\to 0$. The refraction
coefficient is the coefficient $n^2(x)$ in the wave equation
$[\nabla^2+k^2n^2(x)]u=0$.
\end{abstract}
\pn{
 }

\section{Introduction}
The problem we are concerned with is the following: 
how does one  create in a given 
bounded domain $D\subset \R^3$ a material with a desired refraction 
coefficient $n^2(x)$. The domain $D$ originally is assumed to be 
filled with a material with a known refraction coefficient
$n_0^2(x)$. We assume that  Im $n_0^2(x)\geq 0$ and $n_0^2(x)=1$ in 
$D':=\R^3\setminus D$. Originally the wave equation is:
 \be\label{e1}
L_0u_0:=[\nabla^2+k^2n_0^2(x)]u_0=0\quad \text{in }\R^3,\quad
k=const>0, \ee \be\label{e2} u_0=e^{ik\alpha \cdot x}+v_0,\quad
\alpha\in S^2, \ee \be\label{e3}
v_0=A_0(\beta,\alpha,k)\frac{e^{ikr}}{r}+o\left(\frac{1}{r}\right),\quad
r=|x|\to \infty,\quad \beta:=\frac{x}{r}.\ee 
The function $v_0$ is
the scattered field, $A_0(\beta,\alpha,k)$ is the scattering
amplitude, $u(x,\alpha,k)$ is the scattering solution, $S^2$ is the
unit sphere in $\R^3$.\\
We embed $M$ small particles $D_m$, $S_m:=\partial D_m$, $1\leq
m\leq M$, into $D$, so that in any subdomain $\Delta\subset D$ there
are \be\label{e4} \N(\Delta)=\frac{1}{a^{2-\kappa}}\int_\Delta
N(x)dx[1+o(1)],\quad a\to 0 \ee small particles. Here $N(x)\geq 0$
is a continuous (0r piecewise-continuous) function which we can choose
as we wish, $0<\kappa<1$ is a parameter which is our disposal. For
simplicity we assume that particles $D_m$ are balls centered at the
points $x_m$ and of radius $a$ independent of $m$. The distance $d$
between neighboring particles is assumed to be \be\label{e5}
d=O(a^{\frac{2-\kappa}{3}})\quad \text{as }a\to 0. \ee The
properties of a particle are described by the boundary impedance
\be\label{e6} \zeta_m=\frac{h(x_m)}{a^\kappa}, \ee where $h(x)$ is a
continuous function on $D$, Im $h(x)\leq 0$. The function $h(x)$, as
$N(x)$, we can choose as we wish. The scattering solution
$u(x,\alpha,k)$ in the presence of the embedded particles solves the
problem: \be\label{e7} L_0 u=0\quad\text{in
}\R^3\setminus \cup_{m=1}^MD_m,\ee \be\label{e8} 
u_N+\zeta_mu=0\quad\text{on
}S_m,\quad 1\leq m\leq M,\ee \be\label{e9} u=u_0(x,\alpha,k)+v,\ee
\be\label{e10}
v=A_1(\beta,\alpha,k)\frac{e^{ikr}}{r}+o\left(\frac{1}{r}\right),\quad
|x|=r\to \infty,\quad \beta:=\frac{x}{r}.\ee Let us now describe our
results. We prove that problem \eqref{e7}-\eqref{e10} has a unique
solution $u(x,\alpha,k):=u_M(x,\alpha,k)$. We prove that given an
arbitrary function $n^2(x)$ such that $n^2(x)=1$ in $D'$, $n^2(x)$
is continuous or piecewise-continuous in $D$ (with the set of
discontinuities of Lebesgue measure zero in $\R^3$), one can choose
$N(x)$ and $h(x)$ so that the limit 
\be\label{e11}
\psi:=\psi(x,\alpha,k)=\lim_{M\to \infty}u_M(x,\alpha,k)\ee 
exists and
satisfies the equation \be\label{e12}
[\nabla^2+k^2n^2(x)]\psi=0\quad\text{in }\R^3,\ee \be\label{e13}
\psi=u_0(x,\alpha,k)+w(x,\alpha,k),\ee \be\label{e14}
\psi=e^{ik\alpha\cdot
x}+A(\beta,\alpha,k)\frac{e^{ikr}}{r}+o\left(\frac{1}{r}\right),\quad
r=|x|\to \infty,\quad \beta:=\frac{x}{r}.\ee Therefore the medium
with embedded particles in the limit $M\to \infty$, or, which is the
same by \eqref{e4}, in the limit $a\to 0$, has a desired refraction 
coefficient
$n^2(x)$.\\ In Section 2 we formulate the recipe for choosing $N(x)$
and $h(x)$ which guarantees the existence of the limit \eqref{e11}
which solves problem \eqref{e12}-\eqref{e14}. We do not assume that
the small particles are embedded periodically.\\ The aim of this
paper is to make clear for a wide audience of engineers and physicists 
our recipe for creating material with
any desired refraction coefficient and to formulate two
technological problems which must be solved in order that our
theory can be immediately implemented experimentally. Theoretical
justification of our results are given in \cite{R509}-\cite{R536},
see also \cite{R515}- \cite{R476}.

\section{The recipe for creating material with a desired refraction
coefficient} 
The problem we are interested in is the following:

One is given $n_0^2(x)$ and wants to create a
refraction coefficient $n^2(x).$ 

{\it Here is our recipe for doing
this.}\\
\underline{Step 1.} Calculate the
function\be\label{e15}p(x):=k^2[n_0^2(x)-n^2(x)]:=p_1(x)+ip_2(x),\ee
where $p_1(x)=\text{Re }p(x)$, $p_2=\text{Im }p(x)$.\\
\underline{Step 2.} Find two functions $N(x)\geq 0$ and
$h(x)=h_1(x)+ih_2(x)$ from the relation  \be\label{e16} 4\pi
h(x)N(x)=p(x).\ee This can be done by infinitely many ways. For
example, one may fix $N(x)>0$ and define \be\label{e17}
h_1=\frac{p_1(x)}{4\pi N(x)},\quad h_2=\frac{p_2(x)}{4\pi N(x)}. \ee
If one wishes to deal only with passive materials, then one requires
Im $n^2(x)\geq 0$, Im $h(x)\leq 0$, and, if Im $n_0^2(x)\leq$ Im
$n^2(x)$,then Im $p(x)\leq 0$.\\
\underline{Step 3.} Partition the domain $D$ into a union of small
cubes $\Delta_p,$ $1\leq p\leq P,$ without common interior points,
$D=\cup_{p=1}^P\Delta_p$, the center of $\Delta_p$ is denoted by
$y_p$, the side of $\Delta_p$ is of the order
$O(a^{\frac{2-\kappa}{6}}).$ In each cube $\Delta_p$ embed
$\N(\Delta_p)$ small particles, where $\N(\Delta_p)$ is defined in
\eqref{e4}. The distance $d$ between neighboring particles should be
$d=O(a^{\frac{2-\kappa}{3}}).$  The given order of the smallness of $d$
as $a\to 0$ is important, but the distance need not be exactly the 
same. The boundary impedance of each of the
small particle embedded in $\Delta_p$ make equal to
$\frac{h(y_p)}{a^\kappa},$ where $h(x)=h_1(x)+ih_2(x)$ is the function 
found in Step 2 of the recipe.

\begin{thm}\label{thm1}
After the completion of Step 3, the material, obtained from the original 
one with the refraction
coefficient $n_0^2(x)$, will have the refraction coefficient
$n_M^2(x)$, and $\lim_{M\to \infty}n_M^2(x)=n^2(x).$
\end{thm} Proof of this theorem one finds in papers \cite{R509} and
\cite{R536}.
\section{A discussion of the recipe}
Step 1 of the recipe is trivial. Step 2 is also trivial. One may
choose $N(x)>0$ to satisfy some practical requirements. For example,
if one chooses $N(x)$ small, then the total number of particles will
be smaller. Practically one cannot take the limit $M\to \infty$,
i.e., in the limit $a\to 0$, and one stops at some
finite  value of $M$, or of $a>0$.
% What is the relative error $\frac{|n^2_M(x)-n^2(x)|}{n^2(x)}$ in 
%practice an experimental implementation of the recipe should show. 
The two
technological problems, that have to be solved in order that our
recipe can be implemented experimentally, are:\\
1) How does one embed a small particle at a given point into the
given material in $D$?\\
2) How does one prepare a small particle, a ball of radius $a$
centered at a point $x_m$, with the prescribed boundary impedance
$\zeta_m=\frac{h(x_m)}{a^\kappa}$?\\
 Here $h(x)$ is the function,
found at Step 2 of the recipe?

Possibly, the first technological problem can be solved by the
stereolitography process. 

One should be able to solve the second technological problem 
because its limiting cases $\zeta=0$ (acoustically
hard particles, particles from insulating material) and
$\zeta_m=\infty$ (acoustically soft particles, perfectly conducting
particles) are easy to solve in practice, so the intermediate values of 
the boundary impedance should be also possible to prepare.\\
A similar theory has been developed in paper \cite{R535} for
electromagnetic wave scattering by many small dielectric and
conducting particles embedded in an inhomogeneous medium.

\section{Electromagnetic waves}
Assume now that the governing equations are the Maxwell equations
\be\label{e18} \nabla\times E=i\omega \mu H,\quad \nabla\times
H=-i\omega\epsilon'(x)E \quad{\quad in \quad}\R^3,\ee $\mu=const,$
$\epsilon'(x)=\epsilon=const$ in $D'$, $\omega>0$ is frequency,
$\epsilon'(x)=\epsilon(x)+\frac{\sigma(x)}{\omega},\quad
\sigma(x)\geq 0$ is the conductivity, $\sigma(x)=0$ in $D'$. We
assume that $\epsilon'(x)\in C^2(\R^3)$, $\epsilon'(x)\neq 0$, is a
twice continuously differentiable function. Let
$k=\frac{\omega}{c}$, $c=\omega\sqrt{\epsilon \mu}$ is the wave
velocity in $D'$. The incident plane wave is $\Es e^{ik\alpha\cdot
x},\quad \al\in S^2,$ $\alpha\cdot \Es=0,$
$\Es$ is a constant vector.

 Under the above
assumptions the electrical field $E(x)$ is the unique solution to
the equation (see \cite{R535}): \be\label{e19}\begin{split}
E_0(x)&=\Es e^{ik\alpha \cdot x}+ \int_D
g(x,y)p(y)E_0(y)dy\\
&+\nabla_x\int_Dg(x,y)q(y)\cdot E_0(y)dy,\quad
g(x,y):=\frac{e^{ik|x-y|}}{4\pi|x-y|},\end{split}\ee where
\be\label{e20}p(x):=K^2(x)-k^2,\quad K^2(x):=\og^2\ep'(x)\mu;\quad
q(x):=\frac{\nabla K^2(x)}{K^2(x)}. \ee If $M$ small particles
$D_m$, $1\leq m\leq M$, are embedded in $D$, then the basic equation
\eqref{e19} becomes \be\label{e21}
E_M(x)=E_0(x)+\sum_{m=1}^M\int_{D_m}g(x,y)p(y)E_M(y)dy+
\sum_{m=1}^M\int_{D_m}g(x,y)q(y)E_M(y)dy.\ee
It is proved in \cite{R535} that if the size $a$ of small 
particles
tends to zero, if the number of these particles in any open subset
$\Delta$ of $D$ is \be\label{e22}
\N(\Delta)=\frac{1}{a^{3-\kappa}}\int_\Delta N(x)dx[1+o(1)],\quad
a\to 0, \ee and if the distance $d$ between neighboring particles is
$d=O(a^{\frac{3-\kappa}{3}})$, then there exists the limit
$\lim_{M\to 0}E_M(x)=E_e(x)$. 

The limiting field $E_e(x)$, i.e., the
effective field in the medium, solves the equation:
\be\label{e23} E_e(x)=E_0(x)+\int_Dg(x,y)C(y)E_e(y)dy,\quad
C(y)=N(y)c(y), \ee 
where 
\be\label{e24} c(y)=\lim_{a\to 0}
a^{\frac{1}{3-\kappa}}\int_{|y-x|\leq a}p(x)dx.\ee 
If, e.g., the
small particle $D_m$ is a ball of radius $a$ centered at 
a point $y$, and
\bee p(x)=\left\{
        \begin{array}{ll}
          \frac{\gamma(y)}{4\pi a^\kappa}\left(1-\frac{|x|}{a}\right)^2,
 & \hbox{$|x|\leq a $;} \\
          0, & \hbox{$|x|>a$,}
        \end{array}
      \right.
\eee in the coordinate system with the origin at the point $y$, and
$\gamma(y)$ is a number we can choose as we wish, then $c(y)$ in
\eqref{e24} can be calculated: $c(y)=\gamma(y)/30$. Equation
\eqref{e23} implies: 
\be
[\nabla^2+\mathcal{K}^2(x)]E_e=0\quad\text{in }\R^3,\quad
\mathcal{K}^2(x):=K^2+C(x), \ee 
where $C(x)$ is defined in \eqref{e23}. This equation  can be 
rewritten as
\be\label{e26} \nabla\times\nabla\times
E_e=\mathcal{K}^2(x)E_e+\nabla \nabla \cdot E_e. \ee The term
$\nabla \nabla\cdot E_e$ plays the role of the current $i\omega \mu
J$. 

This term can also be interpreted as the term due to a non-local
susceptibility $\chi$: if
$$D_e(x)=\tilde{\ep}(x)E_e-i\og\int_D\chi(x,y)E_e(y)dy,$$ 
then the Maxwell's equations
$$\nabla\times E_e=i\omega \mu H_e,\quad \nabla\times
H_e=-i\omega\tilde{\epsilon(x)}E_e-i\omega\int_D\chi(x,y)E_e(y)dy$$
imply
$$ \nabla\times\nabla\times
E_e=\og^2\tilde{\ep}(x)\mu E_e(x)+\og^2\mu\int_D\chi(x,y)E_e(y)dy.$$
This equation is of the form \eqref{e26} if
$\tilde{\ep}(x)=\frac{\mathcal{K}^2(x)}{\og^2\mu},$ and
$$\chi(x,y)=(\og^2\mu)^{-1}\nabla_x(\dl(x-y)\nabla_y),$$
where $\delta(x-y)$ is the delta-function.

\end{document}